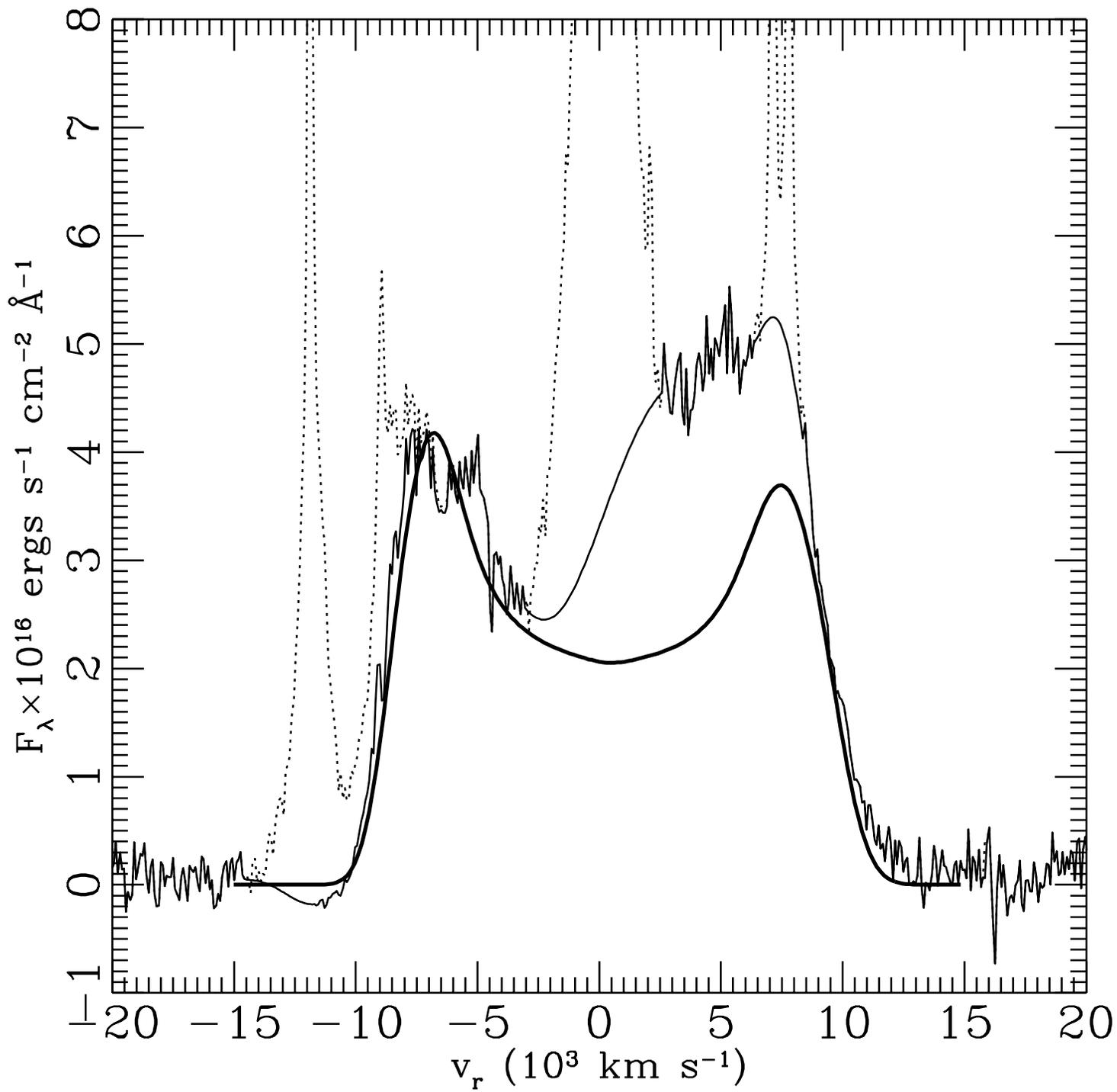

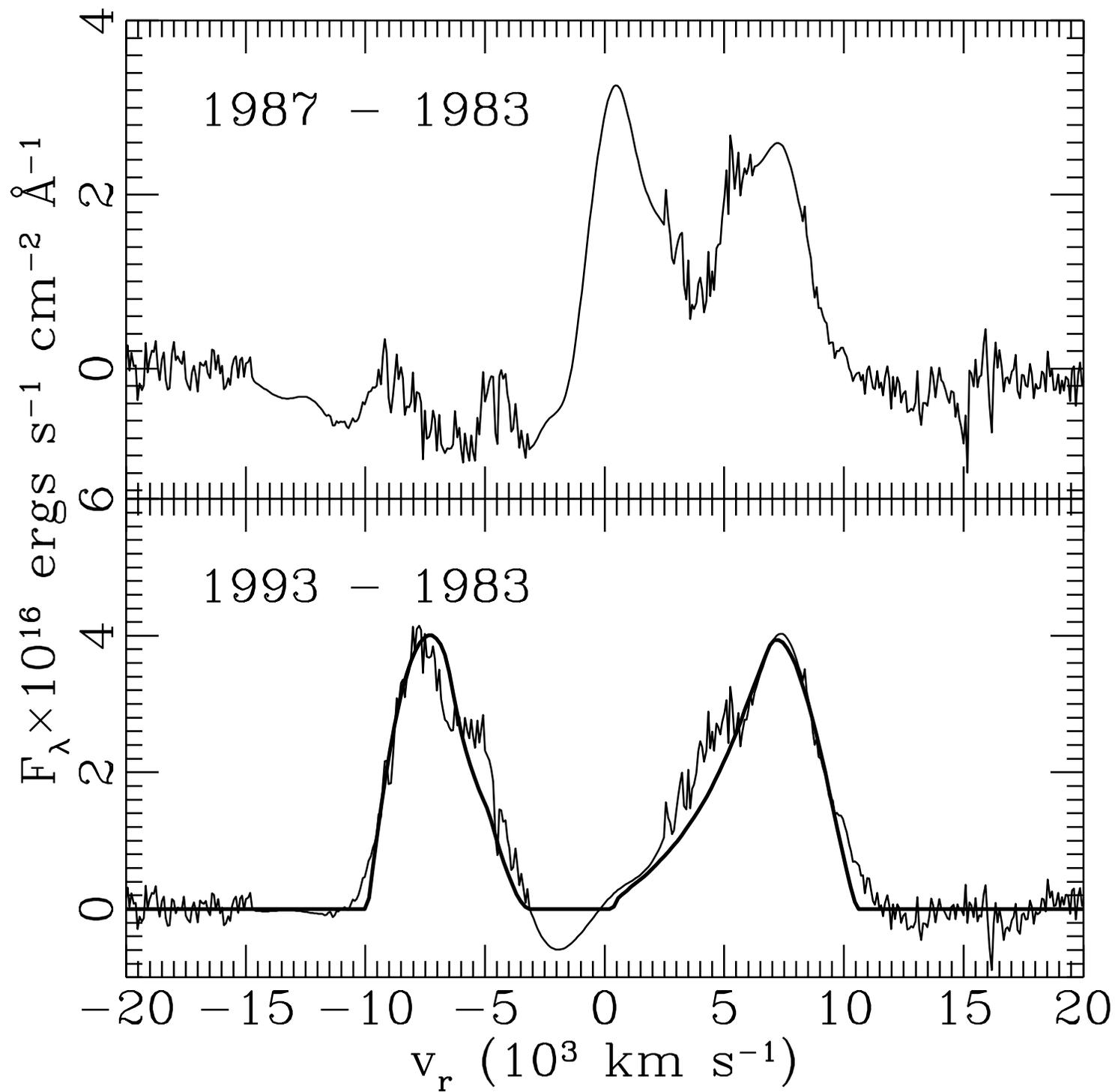

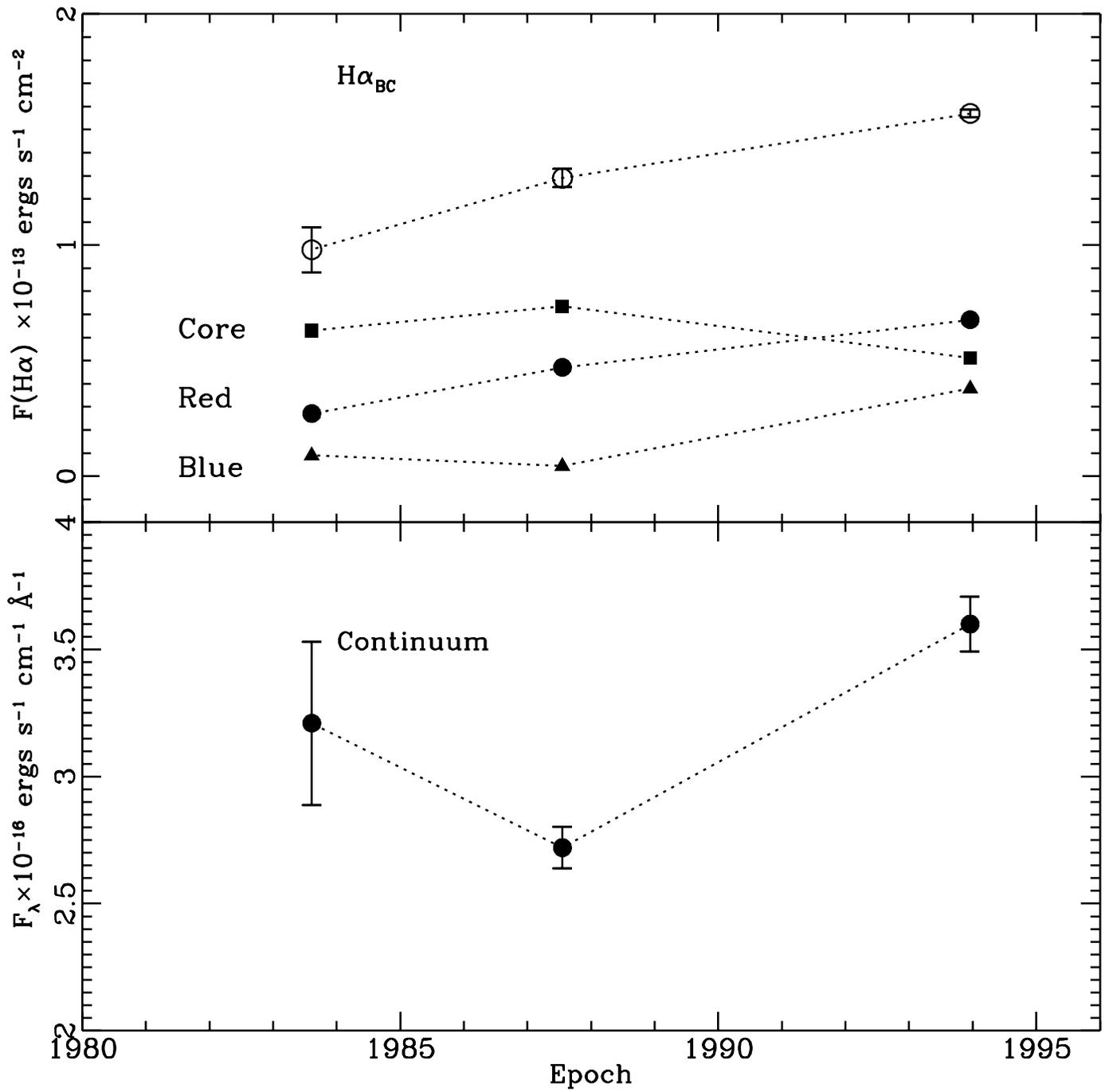

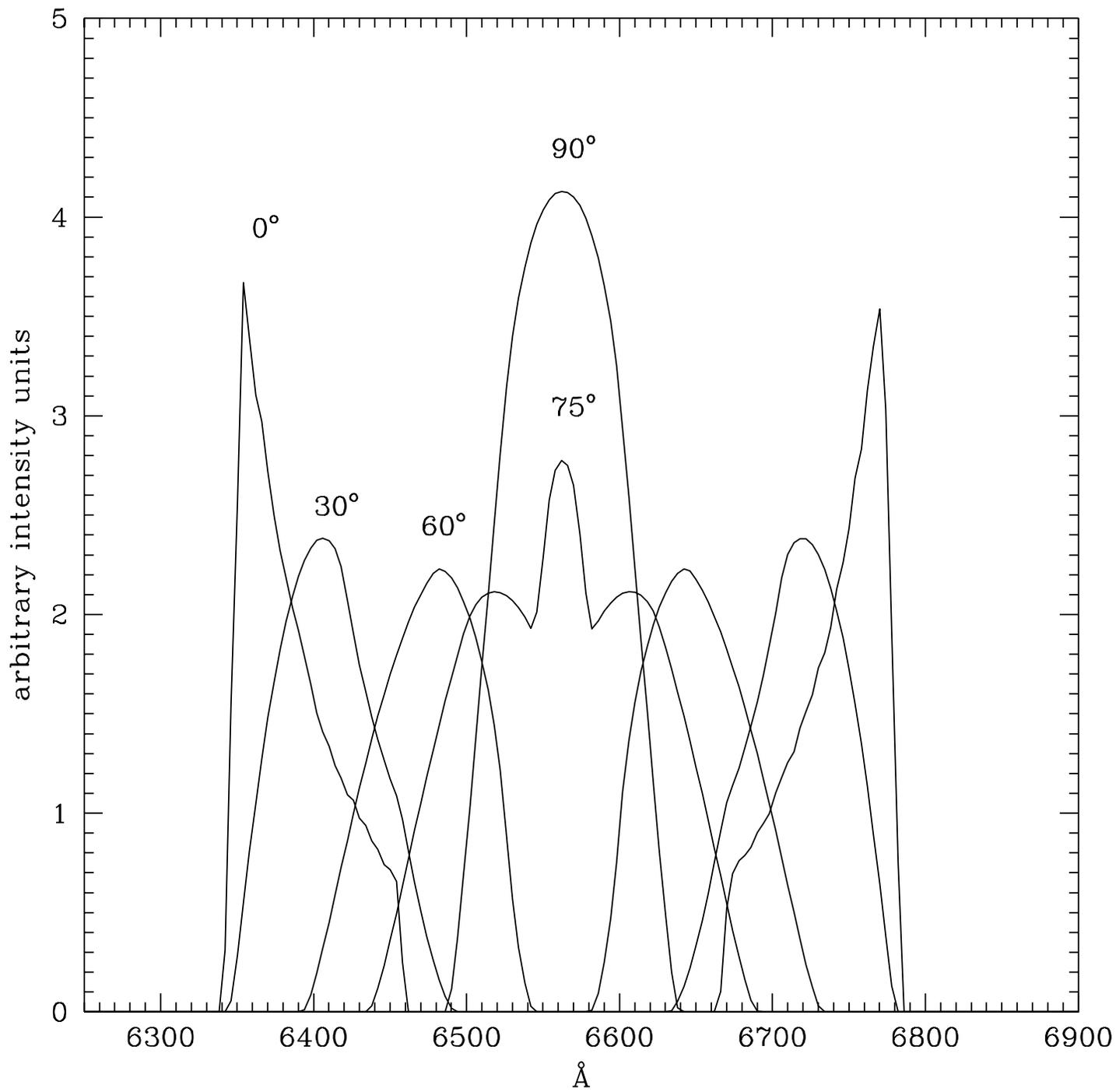

# Pictor A: A New Double-Peaked Emission Line Quasar[1]

J. W. Sulentic & P. Marziani

Department of Physics and Astronomy, University of Alabama, Tuscaloosa USA 35487

T. Zwitter[2]

Dipartimento di Astronomia, Università di Padova, Vicolo dell' Osservatorio 5, I–35122 Padova, Italy

M. Calvani

Osservatorio Astronomico, Vicolo dell'Osservatorio 5, I–35122 Padova, Italy

# ABSTRACT

We obtained a new spectrum for the radio galaxy Pictor A in December 1993. Comparison with previous data obtained in 1983 and 1987 shows that the broad component of the Balmer lines has increased considerably in strength and complexity. The Balmer lines now show a boxy, double peaked profile with FWZI$\sim$24000 km s$^{-1}$. Although the 1993 H$\alpha$ profile is reminiscent of Arp 102B, accretion disk models face immediate difficulty because (1) the redshifted peak in the profile is considerably stronger than the blue one, and (2) the red and blue peaks varied at different epochs. The latter finding strongly favors a predominance of radial motion. We argue that Pictor A, and the rare objects with similar extremely broad emission line profiles, are better understood if the orientation is such that the gas is predominantly moving along the line of sight (i.e. a geometrical extremum).

*Subject headings:* Galaxies: Active – Galaxies: Kinematics & Dynamics – Galaxies: Nuclei – Galaxies: Quasars: Individuals (Pictor A, PKS 0518–45) – Galaxies: Quasars: Emission Lines – Line: Profiles

## 1. Introduction

In 1990 we presented a line profile analysis for 61 active galactic nuclei (AGN) spanning the Seyfert, radio galaxy (BLRG) and quasar phenomena (Sulentic et al. 1990). We compared the low ionization emission line profile shift and asymmetry properties with the predictions of models where the line emission originates from a relativistic accretion disk. This study was in response to the successful fit of an accretion disk model to the Balmer line profiles in Arp 102B (Chen & Halpern 1989). We found a "stochastic" distribution of profile (red and blue) shifts and asymmetries while radiating disk models specifically exclude a wide range of observed profile types. The smoothness (i.e. lack of any bi-modality) of the shift/asymmetry distributions for our sample led us to argue that a disk origin was unlikely for most, if not all, of the low ionization line emission.

Arp 102B shows broad and double-peaked Balmer emission lines. We found no double peaked emission profiles in our sample of 61 AGN (Sulentic 1989). Boroson & Green (1992) also find no profiles of this type in their sample of 87 PG quasars. This suggests that they must be rare. In the past few years considerable effort has been expended to find other Arp 102B–like profiles (Eracleous & Halpern 1994; hereafter EH). A total of twelve cases are now known where the Balmer line profiles are reasonably well fit by an accretion disk model (EH). In view of the fact that these twelve profiles were selected from biased and/or incomplete samples of broad line profiles, this suggests that less than 1% of all AGN show line profiles dominated by a disk-like component (less than 10% of BLRG which are less than 10% of all AGN).

Other exotic classes of profiles were presented by EH including displaced single peaks and double peaked profiles with a stronger red peak. We present here a new example of the latter type with an H$\alpha$ profile with FWZI$\approx$ 24000 km s$^{-1}$. The broad component recently strongly increased in the spectrum of the BLRG Pictor A. Spectra obtained before 1987 (Danziger et al. 1977; Filippenko 1985) showed a more classical BLR with only a weak component of ill-defined broad structure on the red and blue sides of H$\alpha$.

## 2. Observations and Data Analysis

Spectra of Pictor A with a total exposure time of 60 minutes were obtained on December 18, 1993 with the 1.5 m. ESO telescope equipped with a Boller & Chivens spectrograph. A 600 l/mm grating and two arcsec slit width (seeing $\leq$ 1 arcsec) yielded a spectral resolution of 3.5 Å FWHM. The spectrum was reduced with standard IRAF procedures. It was carefully corrected for B–band absorption (affecting the red wing of H$\alpha$) using a model for the feature derived from a standard star observed at a similar zenith distance immediately after the observation of Pictor A. We obtained a S/N$\approx$ 30 in the continuum around 6000 Å. An earlier observation, made available to us by A. Filippenko, was obtained with a B&C spectrograph at the 2.5 m. du Pont telescope at Las Campanas Observatory on August 11, 1983 (details are given in Filippenko 1985). A spectrum was also obtained with the 4 m telescope at CTIO and kindly made available to us by S. Simkin. It is the sum of seven exposures (for a total of 14000 seconds) taken on the nights of July 20-22, 1987 with a 1.8 arcsec slit. The S/N ratio was $\gtrsim$ 50 over the entire H$\alpha$ region.

In order to normalize the 1983, 1987 and 1993 spectra to the same photometric scale the fluxes of [O I] $\lambda$6300 and [S II] $\lambda\lambda$6717, 6730 were used as reference lines. We multiplied the 1987 and 1983 spectra by 1.45 and 0.77$\pm$0.10 respectively (the [O I] $\lambda$6300 and [S II] $\lambda$6717, 6730 required slightly different scale factors in the 1983 spectrum). We assume no change in the semibroad component that dominates [O I] $\lambda$6300 emission because the [O I] $\lambda$6300 profile is very similar in all of the spectra.

H$\alpha$ in the 1993 spectrum is a blend of broad and narrow H$\alpha$, [N II] $\lambda\lambda$6548, 6583 Å, [O I] $\lambda\lambda$6300, 6364, and [S II] $\lambda\lambda$6717, 6730. The uncontaminated broad-line component of H$\alpha$ was obtained by subtracting narrow line H$\alpha$+[NII] $\lambda\lambda$6548, 6583 and [OI] $\lambda\lambda$6300, 6363 using narrow line H$\beta$ as a model template. We note that the template also includes a semibroad component that is clearly seen in the [OI] $\lambda$6300 and H$\beta$ line profiles. The FWZI for this component is remarkably large at $\approx$ 6000 km/s. We removed the narrow line contribution from our 1993 spectrum by first subtracting scaled and shifted H$\beta_{NC}$ profile templates for [OI] $\lambda\lambda$6300, 6363 and H$\alpha$ [with I(H$\alpha$)=3×I(H$\beta$)]. The blend after subtraction of H$\alpha_{NC}$ shows residual [NII] $\lambda\lambda$6548, 6583 also with a semibroad profile. It was therefore similarly modeled and subtracted. Only [SII] $\lambda\lambda$6717, 6730 showed a much narrower profile without semibroad component, in agreement with the correlation between line width and critical density found by Filippenko (1985). Difference profiles (relative to 1993) were obtained by subtracting from the 1983 and 87 spectra the scaled narrow line spectrum extracted from our 1993 data. A similar procedure was used to produce the 1987-83 difference spectrum.

## 3. Results

The deblended H$\alpha$ profile (solid line) and narrow line blends (dotted lines) are shown in Fig. 1. The broad line profile difference spectra for 1987 − 1983 and 1993 − 1983 are shown in the upper and lower panels of Fig. 2 respectively. We summarize here the spectra and the principal changes that were observed over the past ten years.

- The 1983 H$\alpha$ broad line profile was predominantly centrally peaked with FWHM$\approx$9900 km/s. Weak and complex extended structure was observed especially on the red side of the profile.

- The 1987 spectrum shows a strong increase on the red side of the profile as well as an increase in the centrally peaked component.

- The 1993 spectrum shows a roughly box shaped profile with (FWHM≈17750 km/s) and two broad flat topped peaks.

The 1993 – 1983 difference spectrum shows two broad humps that are displaced relative to the systemic velocity by approximately ±7650 km/s. The 1993 – 1987 profile difference further suggests that the red side of Hα remained relatively constant after 1987 while the blue peak appeared for the first time. This suggests that the central component faded between 1987 and 1993. Table 1 summarizes fluxes and line widths for Hα in all spectra as well as Hβ measured in 1993. Fig. 3 shows Hα and continuum flux versus time for broad line Hα and the continuum. Individual flux estimates for the red, central and blue parts of the Hα profile are also plotted. The error estimates shown in Fig. 3 ($3\sigma \sim 10\%$ for 1983 and 3% for 1987 and 93) are derived from the minimum residuals in the scaling of [OI] $\lambda 6300$ and [SII] $\lambda\lambda 6717, 6730$.

### 3.1. Model Fits

It is natural that we would first try to interpret our data in the context of line emitting accretion disk models for the reasons cited in § 1. Our original analysis of Pictor A was made before we were aware of the 1987 spectrum. At that time the difference spectrum in Fig. 2b was the only one available to us. The relativistic accretion disk modeled by Chen & Halpern (1989) has four free parameters: the inclination angle $i$, the inner and outer radii of the line emitting part of the disk $\xi_1$, $\xi_2$ and, as a refinement, a local broadening parameter. The model fit superposed on Fig. 1 was obtained with $i = 60\pm 5°$, $\xi_1= 700\pm25$ and $\xi_2=1500\pm100$ (in units $GM/c^2$, where $M$ is the mass of the black hole). The broadening was found to be quite large corresponding to a velocity dispersion of 850 km s$^{-1}$ (three values were considered 200, 400 and 850 km/s). This is the value that was used for Arp 102B (Chen & Halpern 1989) and could probably be attributed to electron scattering in the atmosphere of the disk. Fig. 1 shows that an emitting disk model fails to reproduce the relative strengths of the red and blue humps, although the fit accounts for the redshifted line profile base.

One problem with the model fit shown in Fig. 1 is a significant residual red bump. Its center is redshifted by $\sim 4500$ km s$^{-1}$ with respect to NLR H$\alpha$. We note that using more symmetrical profiles of the kind predicted by models of Dumont & Collin-Souffrin (1990) cannot resolve the problem of the red bump. The same holds for geometrically thick accretion disks, unless their half thickness is quite large ($\sim 30°$) so that at a given inclination we look just over the surface of the disk. The latter scenario introduces a new free parameter connected with the geometrical shape of the disk. The parameter values for an accretion disk fit in that case are similar to the ones mentioned above except that the the outer radius decreases.

An alternative to disk models involves the possibility that the double peaked profile arises from a biconical structure. The profile shape and variations are then attributed to outflowing gas that is photoionized by the central continuum source. This model has been explored with some success for both double (Zheng, Binette & Sulentic 1990) and single peaked profiles with large radial velocity displacement (Marziani et al. 1993). If we take into account the lack of major changes between 1975 and 1987, it seems more appropriate to model the H$\alpha$ profile as a composite of constant and variable components. Therefore we apply the bicone model to the difference profile 1993 – 1983. It is not appropriate to apply a disk model here because the two peaks vary out of phase.

Fig. 2 shows a bicone fit where disjointed triangular profiles (with the shallowest sides facing each other) naturally occur over a wide range of parameters provided that the bicone is seen at small inclinations (inclination is here defined as the angle between the line of sight and the axis of the cone). The main requirements are: (a) the cone is of moderate aperture (half opening angle 10°≤$\Theta_0$≤30°); (b) the velocity field is in the form

$$v = V_0 \times \sqrt{a + bR_{min}/r},$$

with b<0. This means that the gas is accelerated close to a terminal velocity on a relatively short scale, and (c) that the inclination is $\leq 30°$. The best fit value i≈30° is necessary to explain the width of the two humps (see Fig. 2; parameters are $\Theta_0 = 20°$, $a = 1.00$, $b = -0.90$, $V_0 = 16000$ km s$^{-1}$; $R_{max} = 1.5 R_{min}$ [blue half]; $R_{max} = 1.65\ R_{min}$ [red half]).

## 4. Conclusion

We are able to make only a first order comparison of disk and bicone models in this Letter. However Pictor A only reinforces the already clear problems that exist for accretion disk models: (1) the rarity of very broad profiles, (2) the small number of these profiles that fit the shift and asymmetry properties predicted by the models (see Sulentic et al. 1990) and, for the rare objects with double peaks, (3) the clustering of FWZI and disk inclination around a definite value. Clustering of FWZI seems to suggest a terminal velocity. 8 of the 12 AGN with H$\alpha$ profile reasonably fit by an accretion disk model (EH) have <FWZI> $\approx$ 23500 km s$^{-1}$. Inclination values deduced from best fits are in an intermediate range between 15 and 45°. Accepting the 12 objects found by EH as accretion disk emitters forces one to search for reasons why we do not see more double peaked profiles, with a more uniform distribution of observed and disk model parameters. It is worth noting that 8 out of 12 disk emitter candidates of EH group in the disk model parameter space with $\xi_1$ between 150–450 and $\xi_2$ between 1000–2000.

Bipolar outflows lessen or remove these problems for two reasons: (1) double peaked profiles arise naturally as a relatively rare class because they are viewed only near a geometrical extremum. In Fig. 4 we show the profile dependence on inclination (model parameters left otherwise unchanged from the best fit to the H$\beta$ profile difference for Pic A): wider, more widely disjointed profiles arise if the cone axis is oriented close to the line of sight; narrower, single peaked ones are seen if the cone axis and the line of sight are perpendicular. Since the probability of observing a randomly oriented source is $\propto \sin i$, the widest, double peaked profiles should be of the rarest occurrence, as we indeed observe; (2) good fits can be obtained to both red and blue dominated double peak sources as well as displaced single peaks. A good example of the latter success involves OQ208 (Marziani et al. 1993) which exhibits a single strongly redshifted peak whose variable intensity correlates with peak redshift.

Thus most of the broad profiles with displaced peaks and/or their profile variations can be qualitatively explained with this class of models. The crucial statistical tests for the bicone models will be (1) whether all of the broadest profiles can be fit with a restricted parameter range consistent with pole-on orientation (to avoid the need to explain why more are not observed) and (2) whether the resulting prediction of the number of profiles as a function of FWHM (pure dependence on inclination) will conform to the observations. An outflowing component, superimposed to a profile with monotonically decreasing wings, may explain the larger boxiness of disk–like emitters (EH 12 candidates have mean FWHM/FWZI $= 0.51 \pm 0.10$, while for the other BLRGs it is $0.31 \pm 0.12$; e.g. Miley & Miller 1979, Miller & Peterson 1990).

In the case of Pictor A the scenario that gives the best agreement between data, line profile fits and various heuristic considerations discussed above is one in which the profile variations are due to the symmetric ejection of high velocity gas. We cannot however rule out the possibility of inflow. If the continuum had increased monotonically with time, we could also conclude that the gas is infalling toward the central continuum source, because of the earlier enhancement of the red side. Since the continuum has a minimum in 1987, it is still possible that the blue side of the profile is responding faster to the continuum increase between 1987 and 1993, while the "core" of the broad profile, which is stronger in 1987 (when the continuum was at minimum) might be responding with a larger time delay. We note that the EW of H$\alpha_{BC}$ is consistent with this interpretation. However, more observations of high S/N and resolution at a sampling time shorter than the crossing time of the BLR are needed to distinguish between models invoking inflow and outflow.

Table 1: BROAD LINE PROFILE VARIABILITY

|  | Hα | | | Hβ |
| --- | --- | --- | --- | --- |
|  | 1983 | 1987 | 1993 | 1993 |
| Flux[a] | 0.98 | 1.29 | 1.57 | 0.34 |
| EW (Å) | 260 | 475 | 410 | 82 |
| Continuum[b] | 3.21 | 2.72[c] | 3.60 | 3.94 |
| FWZI (km s$^{-1}$) | 18900 | 18500 | 24400 | 20900 |
| FWHM (km s$^{-1}$) | 8700 | ∼ 7700 | 17400 | ...[d] |

[a]Flux in units of $10^{-13}$ ergs s$^{-1}$ cm$^{-2}$;

[b]Continuum in units of $10^{-16}$ ergs s$^{-1}$ cm$^{-2}$ Å$^{-1}$; averaged on the range 6200 – 6270 Å

[c]Averaged on the range 7050 – 7150 Å.

[d]Ill defined.

# REFERENCES

Boroson, T. & Green, R. 1992, ApJS 80, 109 M. M. 1984, ApJ 286, 464

Chen, K. & Halpern, J. 1989, ApJ 344, 115

Dumont, A.-M., & Collin-Souffrin, S. 1990, A&A 229, 302

Danziger, J., Fosbury, R. & Penston, M. V. 1977, MNRAS 179, 41P

Eracleous, M. & Halpern, J. 1994, ApJS 90, 1 (EH)

Filippenko, A. V. 1985, ApJ 289, 475

Marziani, P., Sulentic, J. W., Calvani, M., Pérez, E., Moles, M. & Penston, M. V. 1993, ApJ 410, 56

Miley, G. K., & Miller, J. S., 1979, ApJ 228, L55

Miller, J. S., & Peterson, B. M., 1990, ApJ 361, 91

Robinson, L., Pérez, E. & Binette, L. 1990, MNRAS 246, 349

Sulentic, J. W. 1989, ApJ 343, 54

Sulentic, J. W. Calvani, M., Marziani, P. & Zheng, W. 1990, ApJ 355, L15

Zheng, W., Binette, L. & Sulentic, J. W. 1990, ApJ 365, 115

We are indebted to A. Filippenko and S. Simkin for the use of the 1983 and 1987 spectra. Extragalactic astronomy at U. Alabama is supported by NSF- EPSCoR RII8996152.

---

This manuscript was prepared with the AAS LaTeX macros v3.0.

Fig. 1.— H$\alpha$ blend as observed in Dec. 1993. The thick line is an accretion disk model fit (see text): the dotted line represents the narrow lines and the solid line shows the pure broad H$\alpha$ component.

Fig. 2.— Upper Panel: H$\alpha$ line profile difference 1987 – 1983; Lower panel: H$\alpha$ line profile difference 1993 – 1983; thick line is "best fit" for an outflowing ensemble of clouds confined in a bi-cone.

Fig. 3.— Continuum (lower panel) and H$\alpha$ flux measurements (upper panel) at different epochs. Open Circles: total H$\alpha_{BC}$ flux; filled circles: red wing of H$\alpha_{BC}$; filled triangles: blue wing; filled squares: core. Core flux was measured in the range $-4400$ km s$^{-1}$ $\lesssim v_r \lesssim$ 2800 km s$^{-1}$.

Fig. 4.— Line profile dependence on inclination (for $i = 0°, 30°, 60°, 75°, 90°$). Horizontal scale is wavelength in Å; vertical scale is arbitrary intensity units.